\documentclass[aps,prc,twocolumn,preprintnumbers,superscriptaddress,nofootinbib]{revtex4-1}
\usepackage[utf8]{inputenc}
\usepackage{graphicx}
\usepackage{psfrag}
\usepackage{amssymb}
\usepackage{amsmath}
\usepackage{epstopdf}
\usepackage{color}
\usepackage{bm}

\begin{document}


\title{Estimation of $\Upsilon$(1S) production in ep process near threshold}

\author{Yu Xu}
\affiliation{Institute of Modern Physics, Chinese Academy of Sciences, Lanzhou 730000, China}
\affiliation{University of Chinese Academy of Sciences, Beijing 100049, China}

\author{Yaping Xie}
\affiliation{Institute of Modern Physics, Chinese Academy of Sciences, Lanzhou 730000, China}
\affiliation{University of Chinese Academy of Sciences, Beijing 100049, China}

\author{Rong Wang}
\email{rwang@impcas.ac.cn}
\affiliation{Institute of Modern Physics, Chinese Academy of Sciences, Lanzhou 730000, China}
\affiliation{University of Chinese Academy of Sciences, Beijing 100049, China}

\author{Xurong Chen}
\email{xchen@impcas.ac.cn}
\affiliation{Institute of Modern Physics, Chinese Academy of Sciences, Lanzhou 730000, China}
\affiliation{University of Chinese Academy of Sciences, Beijing 100049, China}

\date{\today}

\begin{abstract}
The near-threshold photo-productions of heavy quarkonia are important ways
to test the QCD-inspired models and to constrain the gluon distribution of nucleon in the large x region.
Investigating the various models, we choose a photon-gluon fusion model
and a pomeron exchange model for J/$\Psi$ photo-production near threshold,
emphasising on the explanation of the recent experimental measurement by GlueX at JLab.
We find that these two models are not only valid in a wide range of the center-of-mass energy of $\gamma$ and proton,
but also can be generalized to describe the $\Upsilon$(1S) photo-production.
Using these two models, we predict the electro-production cross-sections of $\Upsilon$(1S) at EicC
to be 48 fb to 85 fb at the center-of-mass energy of 16.75 GeV.
\end{abstract}


\maketitle

\section{Introduction}
\label{sec:intro}

The electro-productions of J/$\Psi$ and $\Upsilon$(1S) are among the key reactions
that are going to be measured on electron-ion colliders \cite{Accardi:2012qut,AbelleiraFernandez:2012cc,Chen:2018wyz,Chen:2019equ},
because they are important not only for understanding the interaction mechanism of the heavy quarkonium,
but also for probing the gluonic properties of the nucleon.
These measurements will advance our understanding of quantum chromodynamics (QCD) which governs the
properties of hadrons and the interactions involving hadrons.
The large statistic of the exclusive J/$\Psi$ production data at the hard scale is very helpful
in extracting the generalized parton distribution of the gluon.
The generated heavy quarkonia interact with the intact nucleon,
hence it provide a good opportunity in the study of the bound state between the quarkonium and the nucleon.
Last but not least, the scattering between the heavy quarkonia and the nucleon near threshold
can be used to extract the QCD trace anomaly. The heavy quarkonium photo-production off the nucleon helps us
to probe the nucleon structure, to test QCD dynamics at short distances and to verify QCD inspired models~\cite{Sibirtsev:2004ca}.

There are a lot of experiments that measured the J/$\Psi$ photo-production,
while the photo- and electro-production data of $\Upsilon$(1S) at low energy are much scarce.
Electron ion collider in China (EicC) is optimized to have a 3.5 GeV polarized electron beam
and a 20 GeV polarized proton beam~\cite{Chen:2018wyz,Chen:2019equ}, with the centre-of-mass energy reaching 20 GeV.
EicC will fill up the the absence of experimental measurement of $\Upsilon$(1S) production near threshold.
Hence the main purpose of this work is to check the models of J/$\Psi$ production
and to make a bridge to the prediction of $\Upsilon$(1S) production at EicC.
Recent discovery of the penta-quark $P_c$ make the physicists in predicting the penta-quark $P_b$ in the bottom quark sector,
reasoned from the flavor symmetry of heavy quarks.
The expected penta-quark $P_b$ decays into the $\Upsilon$(1S) and the nucleon.
So, for a better searching of penta-quark state $P_b$, a well understanding of the continuum process
of exclusive $\Upsilon$(1S) production is indispensable.
The differential cross-section of the channel $\gamma p\rightarrow J/\Psi p$ is linked to the QCD trace anomaly $b$~\cite{Kharzeev:1995ij,Kharzeev:1998bz},
and proton mass decomposition.
In principle the study on $\Upsilon$(1S) photo-production near threshold better help us in extracting
the trace anomaly $b$ and in understanding the origins of the proton mass~\cite{Ji:1994av,Ji:1995sv} compared to the J/$\Psi$ channel.

Thanks to the electron accelerators, there are a lot of data available
on the J/$\Psi$ electro- and photo-production off the nucleon.
The data are roughly classified into three energy regions as follows:
high and intermediate energy ZEUS and H1 data~\cite{Chekanov:2002xi,Adloff:2000vm,Aid:1996dn},
low energy Fermilab data~\cite{Binkley:1981kv,Frabetti:1993ux},
and near threshold SLAC and GlueX data~\cite{Camerini:1975cy,Ali:2019lzf}.
How to explain the combined data of a wide energy range especially the recently published GlueX data~\cite{Ali:2019lzf} is a challenge.

There are some empirical formulas to explain the process of $\gamma p\rightarrow J/\Psi p$.
In Ref.~\cite{Brodsky:2000zc}, the two gluon and three gluon exchange models are introduced to describe
the processes of correlated quarks $\gamma qq\rightarrow c\overline{c}qq$ and $\gamma qqq\rightarrow c\overline{c}qqq$ respectively.
This model takes into account the fact that the three target quarks recombine into the final proton
via its form factor after emission of two or three gluons and the coupling of the photon to the $c\overline{c}$ pair.
Since the power term $(1-x)^n$ in this model is valid at $x\simeq 1$
and the form factor of proton is parameterized as $F^2_{g}=exp(1.13t)$
according to the experimental data measured at low beam energies~\cite{Camerini:1975cy,Gittelman:1975ix},
this model is only valid in a limited energy range near threshold.
Besides, a parameterized STZ (Strikman-Tverskoy-Zhalov) model~\cite{Strikman:2005ze} is constructed
to study the incoherent J/$\Psi$ photo-production off a nucleus.
This model shows a good agreement with almost all data due to the application of two heaviside functions.
However such a purely empirical parametrization fitted with the J/$\Psi$ data cannot be applied to the $\Upsilon$(1S) production.

Proton is a complex system of valence quarks, sea quarks and gluons,
hence it is difficult to figure out the explicit process of $\gamma p\rightarrow J/\Psi p$.
Nonetheless, a series of models has been proposed over the passed few decades to explain this process.
Some of them are deduced based on QCD theory and the relevant diagrams.
As gluon is the gauge boson that transmits the strong interaction, in spirit of a factorization theory,
some models exploit the gluon component inside the proton to calculate the heavy vector meson productions.
One of them is the photon-gluon fusion (PGF) model,
which takes into account the photon fusing with a gluon from the proton and transforming into a pair of $c\overline{c}$.
Thus, the cross section is a convolution of the gluon distribution function (GDF)
with the photon-gluon cross section $\sigma_{\gamma g\rightarrow c\overline{c}}$.
On the one side, this model gives a good description of J/$\Psi$ data for almost the whole energy region (see Ref.~\cite{Sibirtsev:2004ca}).
On the other side, this model also provides an experimental approach~\cite{Weiler:1979id} to test the GDF obtained from the global fit
of deep inelastic scattering data, such as the dynamical gluon distribution from IMParton16~\cite{Wang:2016sfq}.
In the dipole model, a colorless object is exchanged between the J/$\Psi$ and the proton.
Since J/$\Psi$ does not couple to a meson and the photon, the pomeron (bunch of gluons) exchange is supposed to be dominant.
Consequently, several pomeron exchange models are proposed to study the photo-production of vector mesons.
Usually, the pomeron exchange model works well at high energy~\cite{Sibirtsev:2004ca}
due to the fact that the gluon distribution is larger than the quark distribution in this region.
However, there are also some empirical pomeron exchange models which coincide well with the experimental data of all energy range.
As $\Upsilon$(1S) production needs high energy, the pomeron exchange model
is appropriate in the study of the photo-production of $\Upsilon$(1S).

In this paper, we use two different models to study the photo-productions of J/$\Psi$ and $\Upsilon$(1S),
which are discussed in Sec.~\ref{sec:JPsiPhotoproduction} and \ref{sec:UpsilonPhotoproduction} respectively.
On this basis, we predict the electro-production of $\Upsilon$(1S) at EicC in Sec.~\ref{sec:UpsilonElectroproduction}.
Lastly some discussions and a summary is given in Sec. \ref{sec:summary}.

\section{Photo-production of J/$\Psi$}
\label{sec:JPsiPhotoproduction}

Different model on the cross-section of $\gamma p\rightarrow J/\Psi p$ has its own applicable energy region
due to its specific physical mechanism.
As discussed in the previous section, we notice that the PGF model and the pomeron exchange model
are in good agreement with the old $J/\Psi$ production data at both high and low energy.
If they are also consistent with the recent published data by GlueX and SLAC \cite{Ali:2019lzf},
then they are suitable to reproduce the heavy quarkonia production from the threshold to a relatively high energy.
As a consequence, these models are then thought to be valid to fit the $\Upsilon$(1S) production
data at high energy and to extrapolate the cross-sections to a lower energy, even near the production threshold.
The reason is that both J/$\Psi$ and $\Upsilon$(1S) have the heavy constituents moving slowly inside
and are of the same quantum numbers.

\subsection{Photon-gluon fusion model}
\label{sec:PGF-model}

\begin{figure}[htbp]
\centering
\includegraphics[width=.32\textwidth]{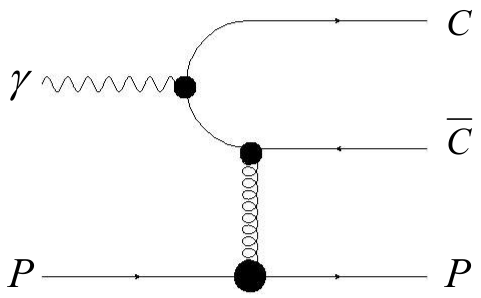}
\caption{The schematic diagram of the photon-gluon fusion model.}
\label{fig:PGF-model}
\end{figure}

In the photon-gluon fusion model, the photon fuses with a gluon emitted from the proton
and split into a $c\overline{c}$ pair, which is shown in Fig. \ref{fig:PGF-model}.
The cross section of charmonium photo-production is given by~\cite{Sibirtsev:2004ca},
\begin{equation}
\sigma_{\gamma p\to pJ/\Psi}=f\int^{4m^2_D}_{4m^2_c}\frac{d\overline{s}}{\overline{s}}
\sigma_{\gamma g\rightarrow c\overline{c}}g(x=\frac{\overline{s}}{s},m^2_{J/\Psi}),
\end{equation}
where $\sigma_{\gamma g\rightarrow c\overline{c}}$ denotes the photon-gluon fusion cross section,
$g(x,m^2_{J/\Psi})$ is the gluon distribution of the proton at $\mu^2=m^2_{J/\Psi}$,
and $f$ is the probability of the $c\overline{c}$ pair color-neutralized into the J/$\Psi$ meson.
$\sigma_{\gamma g\rightarrow c\overline{c}}$ takes the form as~\cite{Weiler:1979id},
\begin{equation}
\begin{aligned}
\sigma_{\gamma g\rightarrow c\overline{c}}=&\frac{2\pi\alpha e^2_c\alpha_s}{\overline{s}^3}\{[\overline{s}^2+4m^2_c(\overline{s}^2-2m^2_c)] \\
&\times ln\left(\frac{1+\beta}{1-\beta}\right)-\beta[\overline{s}^2+4\overline{s}m^2_c]\},
\end{aligned}
\end{equation}
in which $\alpha=1/137$ is the fine structure constant,
$e_c$ and $m_c$ are the charge and the mass of charm quark respectively,
$\alpha_s=0.5$ is the strong coupling constant at $\mu^2=m^2_{J/\Psi}$,
$\overline{s}$ is the squared invariant mass of the photon-gluon system,
and $\beta$ is defined as $\beta^2=1-\frac{4m^2_c}{\overline{s}}$~\cite{Sibirtsev:2004ca}.
The integration range over $\overline{s}$ is from the charmonium production threshold $4m_c^2$
to the open charm production threshold $4m^2_D$. Once above the $D\bar{D}$ threshold, the energetic
$c\overline{c}$ from the photon-gluon fusion forms into D mesons.
In this work, we use the dynamical gluon distribution from IMParton16~\cite{Wang:2016sfq}.

Because one gluon carries a color octet, the $c\overline{c}$ pair must radiate one or more soft gluons
before the hadronization into a color singlet state of J/$\Psi$ wave function.
Yet the PGF model does not implement explicitly this color-neutralization process,
an adjustable parameter $f$ is used to describe the probability of the corresponding hadronization.
The parameter $f$ is energy-dependent and goes down as the energy goes up.
Therefore we apply a power function to describe the energy-dependence as $f=f_0 W^\lambda$.
Performing a fit to the experimental data~\cite{Chekanov:2002xi,Binkley:1981kv,Frabetti:1993ux,Ali:2019lzf},
we get $f_0=7.36$ and $\lambda=-3.08$.
As showed in Fig.~\ref{fig:jpsi-xsection-fits}, the PGF model with the dynamical gluon distribution
is consistent with the experimental data of $J/\Psi$ photo-production.

\subsection{Pomeron exchange model}
\label{sec:pomeron-model}

\begin{figure}[htbp]
\centering
\includegraphics[width=.35\textwidth]{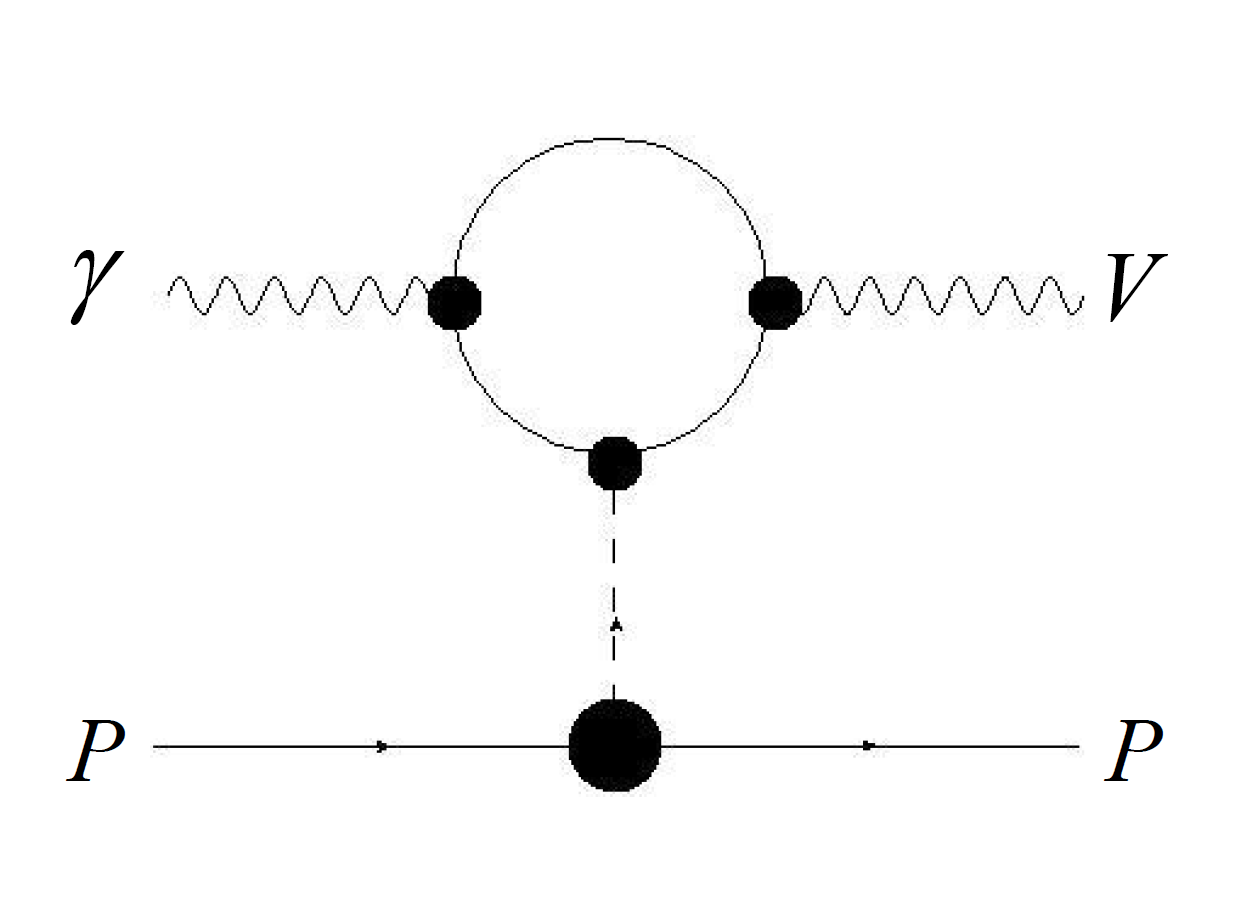}
\caption{The schematic diagram of the pomeron exchange model.}
\label{fig:pomeron-model}
\end{figure}

The phenomenological pomeron exchange model being considered here is
the one proposed in Ref.~\cite{Klein:2016yzr} to describe the cross section of $\gamma p\rightarrow V p$.
The schematic Feynman diagram of the model is shown in Fig. \ref{fig:pomeron-model}.
The cross section of $\gamma p\rightarrow V p$ is parameterized as,
\begin{equation}
\sigma(\gamma p\rightarrow Vp)=\sigma_p\left[1-\frac{(m_p+m_V)^2}{W^2_{\gamma p}}\right]^2W^\epsilon_{\gamma p},
\end{equation}
where the square factor is derived from the phase space and it decreases to zero at the threshold.
The free parameters $\sigma_p$ and $\epsilon$ are determined to be 17.1 nb and 0.109 respectively,
from a fit to the experimental data~\cite{Chekanov:2002xi,Binkley:1981kv,Frabetti:1993ux,Ali:2019lzf}.

\begin{figure}[htbp]
\centering
\includegraphics[width=.45\textwidth]{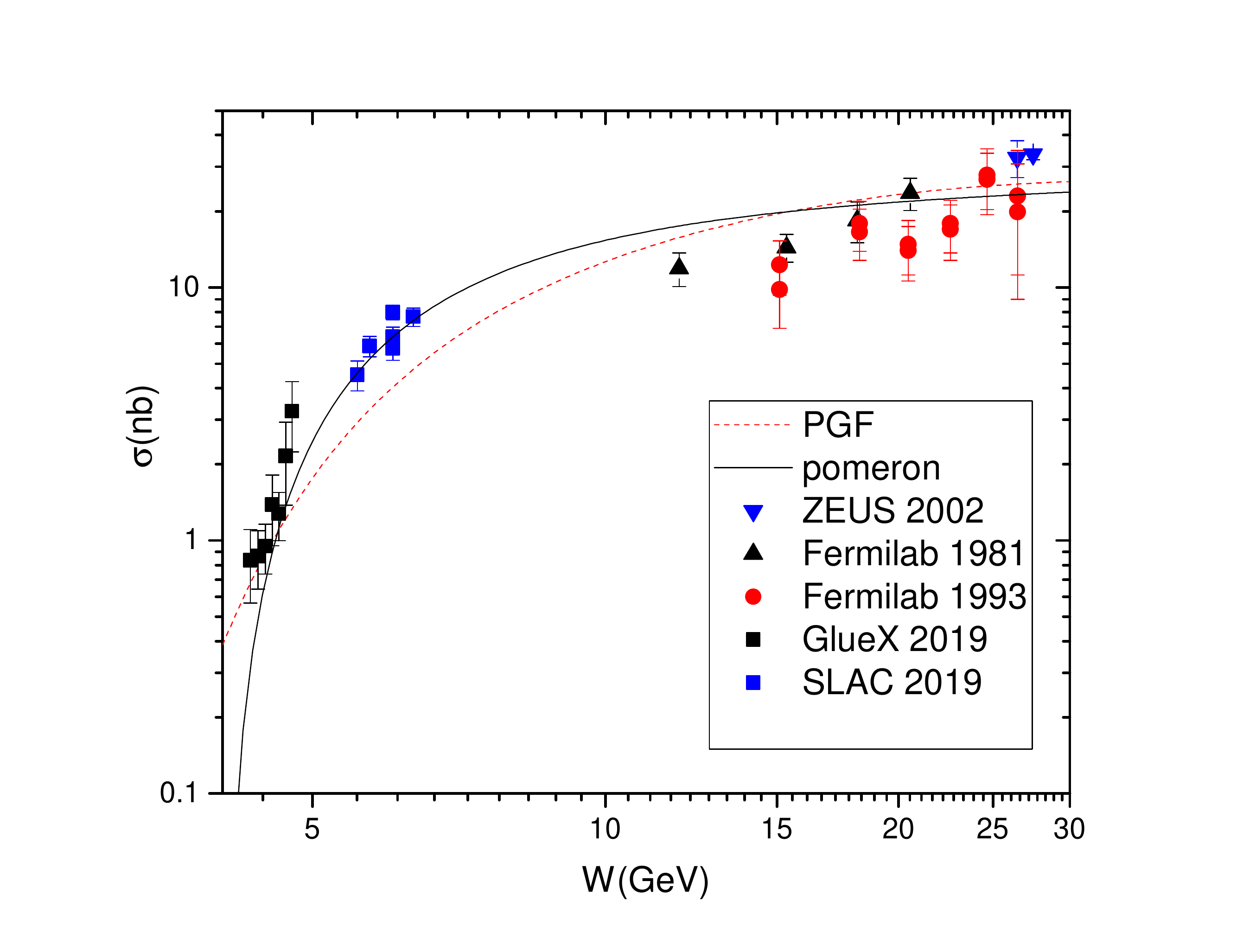}
\caption{The total cross section of $\gamma p\rightarrow J/\Psi p$.}
\label{fig:jpsi-xsection-fits}
\end{figure}

Fig. \ref{fig:jpsi-xsection-fits} shows the comparisons of the pomeron exchange model prediction
with the experimental data and the PGF model prediction.
The total cross-section data of SLAC are deduced from the differential cross-section data \cite{Ali:2019lzf}.
Both models give good explanations for the J/$\Psi$ photo-production near the threshold.

\section{Photo-production of $\Upsilon$(1S)}
\label{sec:UpsilonPhotoproduction}

We use the same theoretical frameworks discussed in the previous section
to predict the photo-production of $\Upsilon$(1S), for they are successful.
Parameters related to the charm quark and J/$\Psi$ should be replaced with the bottom quark and $\Upsilon$(1S).
In the PGF model, the charge and the mass of bottom quark are 1/3 and 4.67 GeV respectively.
The mass of $\Upsilon$(1S) is 9.46 GeV, and the strong coupling constant is taken as $\alpha_s=0.2$ at this mass scale.
The GDF for this channel is at a higher scale as $g(x,m_\Upsilon^2)$.
The upper limit of the integral in PGF model is $4m^2_B$ accordingly.
Then, the free parameters used to describe the hadronization into $\Upsilon$(1S) are obtained to be $f_0=0.328$ and $\lambda=-2.15$,
from a fit to the experimental data~\cite{Aaij:2015kea,Breitweg:1998ki,Chekanov:2009zz,Adloff:2000vm,CMS:2016nct}.
For the pomeron exchange model, the free parameters are obtained to be $\sigma_p=5.19$ pb and $\epsilon=0.787$,
which agrees with a previous study~\cite{Klein:2016yzr}.

\begin{figure}[htbp]
\centering
\includegraphics[width=.45\textwidth]{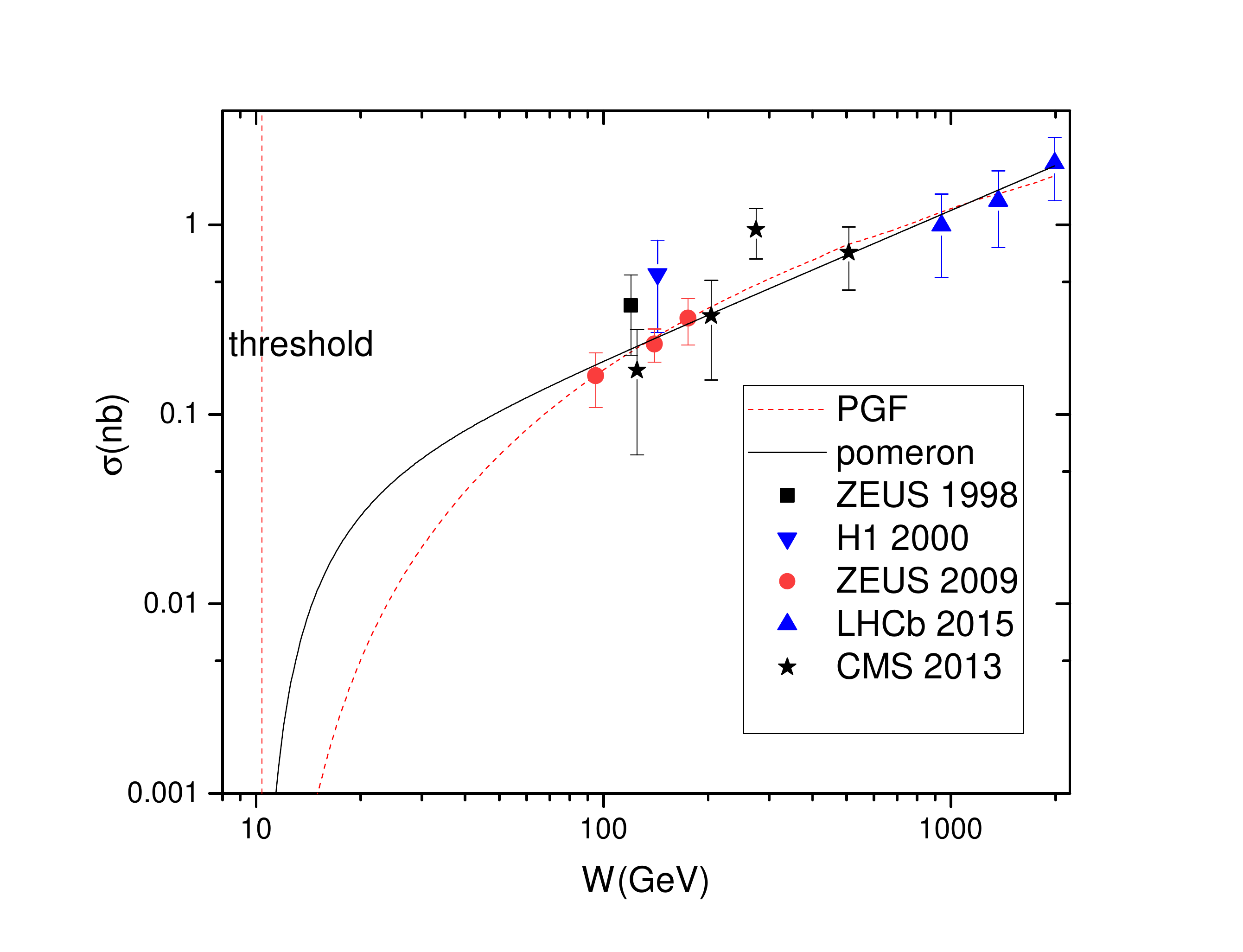}
\caption{The total cross section of $\gamma p\rightarrow \Upsilon p$.}
\label{fig:upsilon-xsection-fits}
\end{figure}

The comparisons of the model calculations with the experimental data on the $\Upsilon$(1S) photo-production
are shown in Fig.~\ref{fig:upsilon-xsection-fits}. Both models are consistent with the existing data amazingly.
The qualities of the fits are $\chi^2 / N=0.46$ and $\chi^2 / N=0.51$
for the photon-gluon fusion model and the pomeron exchange model respectively.
Therefore, we can use these two models to predict the electro-production of $\Upsilon$(1S) at EicC.

\section{Electro-production of J/$\Psi$ and $\Upsilon$(1S)}
\label{sec:UpsilonElectroproduction}

The cross section of $ep\rightarrow epV$ is expressed as,
\begin{equation}
\begin{aligned}
\sigma(ep\rightarrow epV)={}&\int dk\int dQ^2\frac{d^2N_\gamma}{dkdQ^2}{}\\&\times\sigma_{\gamma^*p\rightarrow Vp}(W,Q^2),
\end{aligned}
\end{equation}
where the photon flux~\cite{Budnev:1974de} is,
\begin{equation}
\begin{aligned}
\frac{d^2N_\gamma}{dkdQ^2}=&
\frac{\alpha}{\pi k Q^2}\\
&\times \left[1-\frac{k}{E_e}+\frac{k^2}{2E^2_e}-\left(1-\frac{k}{E_e}\right)\frac{Q^2_{min}}{Q^2}\right],
\end{aligned}
\end{equation}
which represents the probability of electron emitting the virtual photons.
$k$ denotes the energy of the virtual photon, $Q^2$ denotes the minus four momentum square of it
and the minimum $Q^2$ is evaluated as $Q^2_{min}=m^2_ek^2/[E_e(E_e-k)]$.
Note that the photon flux from the electron beam is defined in the proton rest frame.
The vector meson production by the virtual photon is $Q^2$-dependent,
and it is connected with the production by the real photon as~\cite{Adloff:1999kg},
\begin{equation}
\begin{aligned}
\sigma_{\gamma^*p\rightarrow Vp}(W,Q^2)={}&f(M_V)\sigma(W,Q^2=0){}
\\&\times\left(\frac{M^2_V}{M^2_V+Q^2}\right)^n,
\end{aligned}
\end{equation}
where $f(M_V)$ is a Breit-Wigner function~\cite{Klein:1999qj}
and $n=c_1+c_2(Q^2+M^2_V)$, with $c_1=2.09$ and $c_2=0.0073$ determined by the experimental data~\cite{Lomnitz:2018juf}.
The integration over the virtual photon energy $k$ ranges from the threshold of $\Upsilon$(1S) production
to the maximum $k_{max}=E_e-10m_e$.
Since the enormous vector meson productions are at the low virtuality $Q^2$,
the integration over $Q^2$ is taken as $0<Q^2<1$ GeV$^2$ in the calculation.
Using the formulas mentioned above, we finally obtain the electro-production cross section
of $\Upsilon(1S)$ at the optimal energy of EicC, which is 48 fb for the PGF model
and 85 fb for the pomeron exchange model.
As the luminosity of EicC is about $4\times 10^{33}$ cm$^{-2}$s$^{-1}$,
its integral luminosity is around 50 fb$^{-1}$ with a year of operation~\cite{Chen:2018wyz,Chen:2019equ}.
Assuming a detection efficiency of 0.2, there are about 480 to 850 $\Upsilon$(1S) events
which can be collected per year at EicC.

\section{Discussions and summary}
\label{sec:summary}

We investigate two models on the cross-sections of $\gamma p\rightarrow J/\psi p$ and $\gamma p\rightarrow \Upsilon p$.
The photo-gluon fusion model is on the basis of parton degree of freedom, and the validity
of the factorization framework. Our result shows that both the dynamical gluon distribution
and the factorization assumption work fine for a wide range of energy of the system.
This is due the heavy quark that is connected with the gluon is large ($>1$ GeV).
For the pomeron exchange model, the colorless component is exchanged.
Single gluon information is embedded in the coherent soft gluons.
Though it is not related to the parton picture of the hadron,
the pomeron exchange model explains well the experimental data with a few parameters.
According to Ref.~\cite{Sibirtsev:2004ca,Klein:2016yzr}, the two models in this work
also work successfully in explaining the J/$\psi$ photo-production at much higher energy.
The pomeron exchange model is a little better in explaining the J/$\Psi$ near-threshold production
compared to the PGF model, and it is close to the result based on pomeron
trajectory in Ref.~\cite{Cao:2019gqo}.
Although there are just a little $\Upsilon$(1S) photo-production data in the low energy region,
both models in this work match the data successfully.

Electro-production cross-sections of $\Upsilon$(1S) are calculated with the photo-production cross-sections
based on the PGF model and the pomeron exchange model.
The extrapolations of the near-threshold cross-sections from the two models exhibit some differences,
however they are at the same magnitude.
Finally we provide a prediction of the number of $\Upsilon$(1S) collected at EicC
near the threshold with the overall detector efficiency of 20\%,
which is around 500 events per year at the center-of-mass energy of 16.75 GeV.
With years of collections of the $\Upsilon$(1S) events at several energy points near the threshold,
the EicC machine will advance our knowledge about the QCD trace anomaly, the proton mass decomposition,
and the exotic penta-quark baryon in the bottom quark sector.

\begin{acknowledgments}
Sincere acknowledgment is given to Dr. Xu CAO for the useful discussions with him and his enlightening suggestions.
This work is supported by the Key Research Program of the Chinese Academy of Sciences under the Grant NO. XDPB09.
\end{acknowledgments}

\bibliographystyle{apsrev4-1}
\bibliography{refs}
\end{document}